\documentclass[vecphys]{svmult}

\usepackage{makeidx}         
\usepackage{graphicx}        
\usepackage{multicol}        
\usepackage[bottom]{footmisc}

\makeindex             

\begin{document}

\newcommand{\za}{z_0}
\newcommand{\rd}{{\rm d}}
\newcommand{\beq}{\begin{equation}}
\newcommand{\eeq}{\end{equation}}
\newcommand{\bea}{\begin{eqnarray}}
\newcommand{\eea}{\end{eqnarray}}
\newcommand{\parr}{\parallel}
\newcommand{\pa}{\partial}
\newcommand{\cs}{c_{\rm s}}
\newcommand{\va}{v_{\rm A}}
\newcommand{\apj}{ApJ, }
\newcommand{\apjl}{ApJ, }
\newcommand{\nat}{Nature, }
\newcommand{\apss}{Ap.\ Space Sci., }
\newcommand{\aap}{A\&A, }
\newcommand{\pasj}{PASJ, }
\newcommand{\jgr}{J.\ Geophys.\ Res., }
\newcommand{\mnras}{MNRAS, }

\title*{Theory of magnetically powered jets}
\author{H.C.\ Spruit\inst{}}
\institute{Max-Planck-Instititut f\"ur Astrophysik, Postfach 1317, D-85741 Garching 
\texttt{henk@mpa-garching.mpg.de}}

\maketitle

\abstract{The magnetic theory for the production of jets by accreting objects is reviewed with emphasis on outstanding problem areas. An effort is made to show the connections behind the occasionally diverging nomenclature in the literature, to contrast the different points of view about basic mechanisms, and to highlight concepts for interpreting the results of numerical simulations.  The role of dissipation of magnetic energy in accelerating the flow is discussed, and its importance for explaining high Lorentz factors. The collimation of jets to the observed narrow angles is discussed, including a critical discussion of the role of  `hoop stress'. The transition between disk and outflow is one of the least understood parts of the magnetic theory; its role in setting the mass flux in the wind, in possible modulations of the mass flux, and the uncertainties in treating it realistically are discussed. Current views on most of these problems are still strongly influenced by the restriction to 2 dimensions (axisymmetry) in previous analytical and numerical work; 3-D effects likely to be important are suggested. An interesting problem area is the nature and origin of the strong, preferably highly ordered magnetic fields known to work best for jet production. The observational evidence for such fields and their behavior in numerical simulations is discussed. I argue that the presence or absence of such fields may well be the `second parameter' governing not only the presence of jets but also the X-ray spectra and timing behavior of X-ray binaries.}

\section{The standard magnetic acceleration model}
The magnetic model has become the de facto standard for explaining (relativistic) jets, that is,
collimated outflows. In part this has been a process of elimination of alternatives, in part it is
due to analytic and numerical work which has provided a sound theoretical basis for some essential 
aspects of the mechanism. It should be remembered that a key observational test of the model is still
largely missing. Evidence for magnetic fields of the configuration and strength required by the
model is indirect at best. Magnetic fields are detected indirectly through synchrotron radiation
(such as the radio emission of extragalactic jets), and in some cases directly through the Zeeman
effect in spectral lines (OH or H$_2$O masers) in young stellar objects and protoplanetary nebulae
(e.g. \cite{hutawara,bains,vlemmings}). Most of these detections, however, do not refer to the inner regions of the flow where much of the magnetic action is expected to take place. On the theoretical side, the acceleration process itself is the best studied aspect. Problems such as the precise conditions leading to the launching of a flow from  a magnetic object, or the collimation of the flow to a jet-like state are still under debate.

In the magnetic model, outflows are produced by magnetic fields of a (rapidly) rotating object.  These objects include rapidly rotating magnetically active stars, young pulsars, or accretion disks such as those in young stellar objects, X-ray binaries, Cataclysmic Variables and active galactic nuclei (AGN). In these cases, the magnetic fields are `anchored' in the material of the rotating object. A related kind of process is the Blandford-Znajek mechanism \cite{BZ}, in which a rotating black hole with an externally imposed magnetic field is the energy source of a flow.  In the following I limit the discussion to the illustrative case of flows from accretion disks, for which much observational data is available. For more on the Blandford-Znajek mechanism see \cite{komissarov}

\subsubsection{Flow regions}
In the standard magneto-centrifugal acceleration model for jets produced by an accretion disk \cite{kogan76,blandfordpayne} there are three distinct regions. The first is the accretion disk; here the kinetic energy of rotation (perhaps also the gas pressure) dominates over the magnetic energy density. As a result, the field lines corotate with the disk in this region: they are `anchored' in the disk.

\begin{figure}
\centering
\includegraphics[height=6cm]{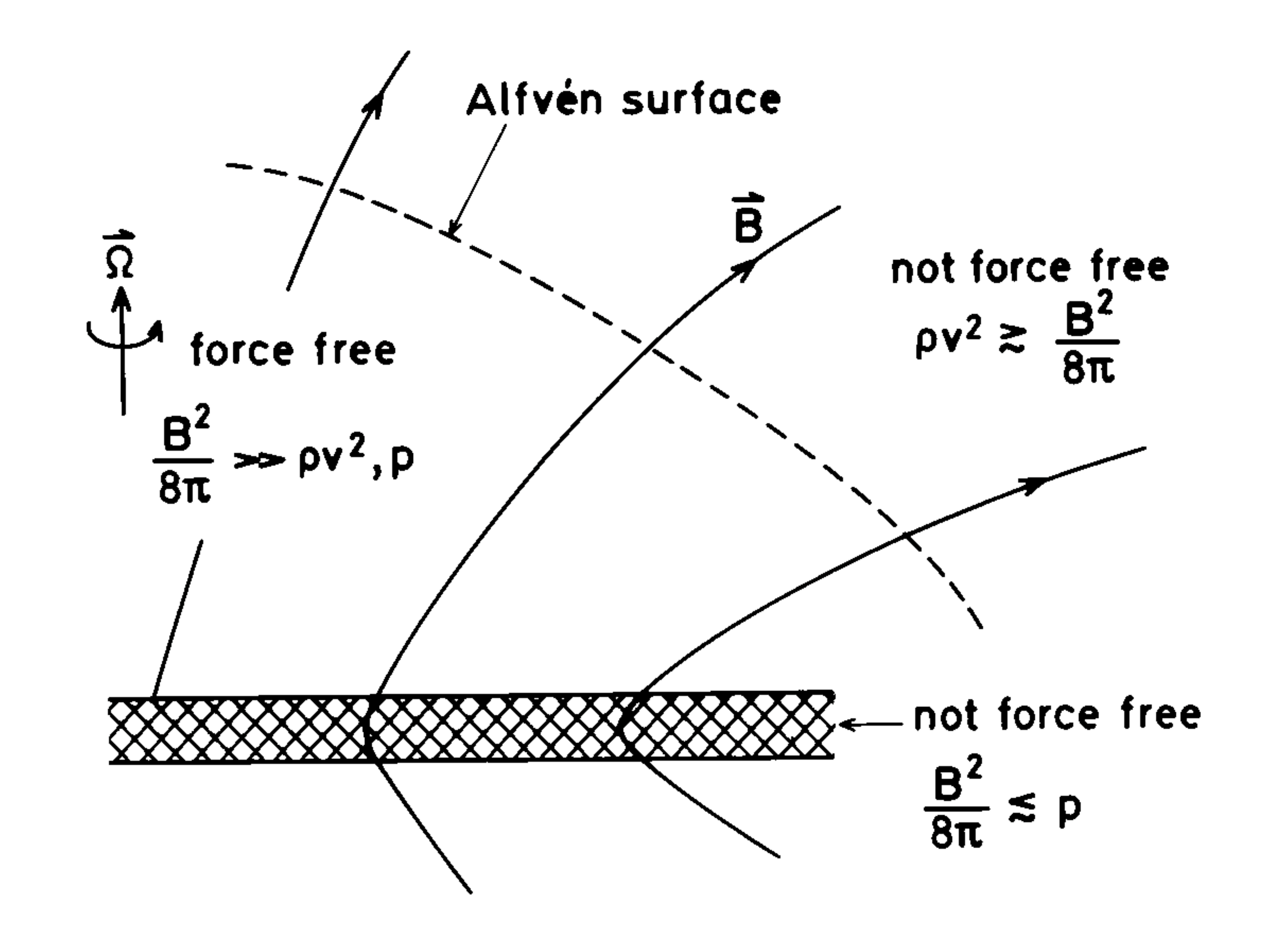}
\caption{Regions in a magnetically accelerated flow from an accretion disk (central object is assumed at the left of the sketch). In the atmosphere of the disk up to the Alfv\'en surface the magnetic field dominates over gas pressure and kinetic energy of the flow. This is the region of centrifugal acceleration.}
\label{regions}     
\end{figure}

The second is a region extending above and below the disk. Assuming the disk to be cool, the atmosphere of the disk has a low density and gas pressure. In this region, the magnetic pressure dominates over gas pressure, so that the field must be approximately force-free [$(\nabla\times{\bf B})\times{\bf B}=0$], like the magnetic field in much of the solar atmosphere. It forces the flow of gas into corotation with the disk, with only the velocity component along the field unrestricted by magnetic forces. The flow experiences a centrifugal force accelerating it along the field lines, much as if it were carried in a set of rotating rigid tubes anchored in the disk. 

This acceleration depends on the inclination of the field lines: there is a net upward force along the field lines only if they are inclined outward at a sufficient angle. Field lines more parallel to the axis do not accelerate a flow. The conditions for collimation and acceleration thus conflict somewhat with each other. Explanation of the very high degree of collimation observed in some jets thus requires additional arguments, in the magnetic acceleration model (see Sect. \ref{colli}).

\begin{figure}
\centering
\includegraphics[height=6cm]{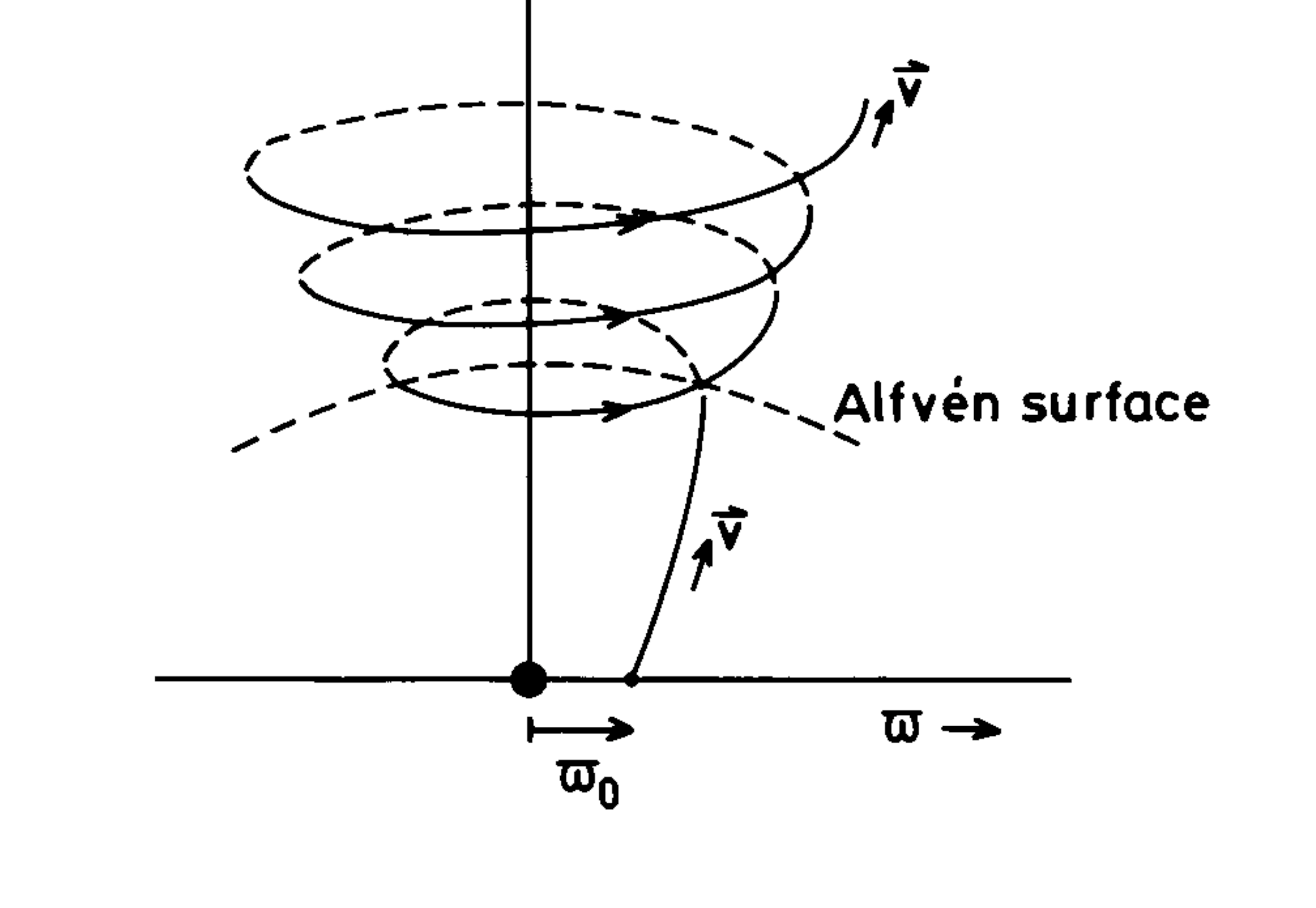}
\caption{Beyond the Alfv\'en distance the field lines lag behind the rotation of their footpoints and are coiled into a spiral (very schematic: the Alfv\'en surface actually has a more complicated shape).}
\label{windup}     
\end{figure}

Finally, as the flow accelerates and the field strength decreases with distance from the disk, the approximation of rigid corotation of the gas with the field lines stops being valid. This happens roughly at the {\em Alfv\'en radius}: the point where the flow speed equals the Alfv\'en speed (for exceptions see Sect. \ref{hiflo}). At this point, the flow has reached a significant fraction of its terminal value. The field lines start lagging behind, with the consequence that they get `wound up' into a spiral. Beyond the Alfv\'en radius, the rotation rate of the flow gradually vanishes by the tendency to conserve angular momentum, as the flow continues to expands away from the axis. If nothing else were happening, the field in this region would thus be almost purely azimuthal, with one loop of azimuthal field being added to the flow for each orbit of the anchoring point, see Fig.\ \ref{windup}. In fact, this state is not likely to survive for much of a distance beyond $r_{\rm A}$, because of other (3-dimensional) things actually happening (see Sect. \ref{dissip}).

\begin{figure}
\centering
\includegraphics[width=\hsize]{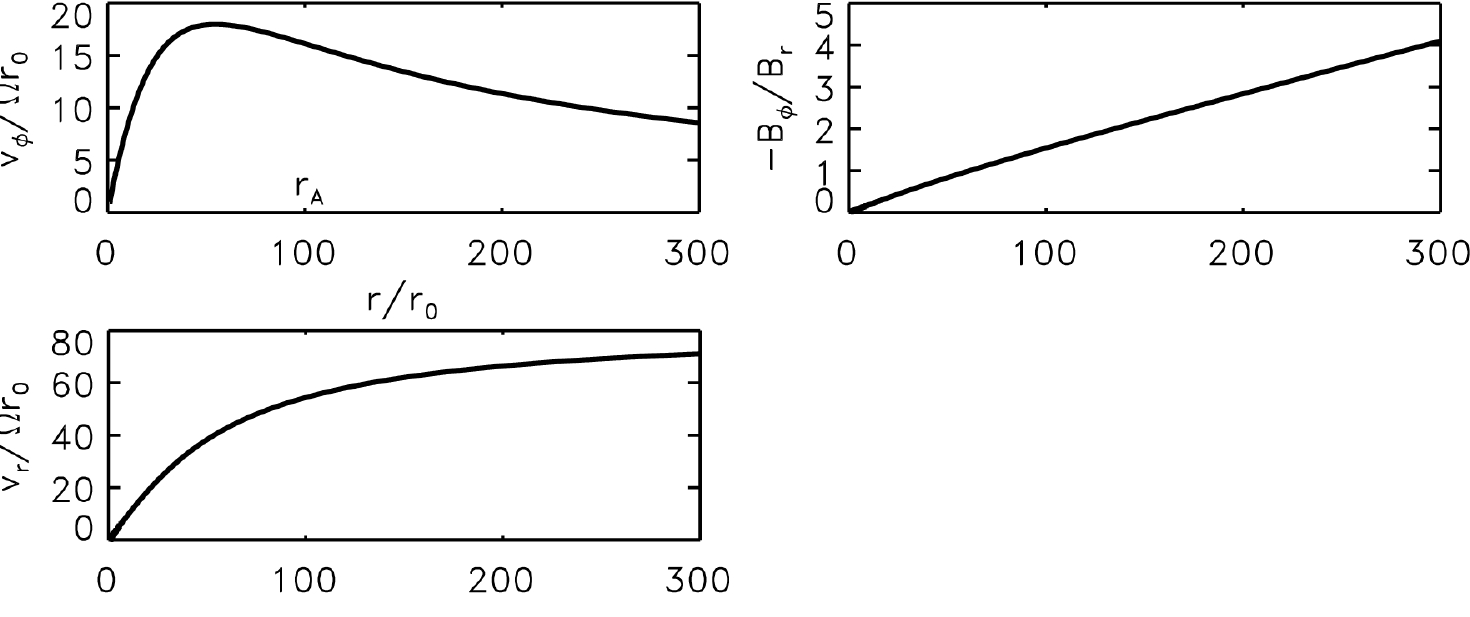}
\caption{Properties of a magneto-centrifugally accelerated flow. Rotation rate, flow speed and azimuthal field angle as functions of distance from the rotation axis  (cold Weber-Davis model). The Alfv\'en distance in this example is at 100 times the footpoint distance $r_0$. }
\label{propts}     
\end{figure}

\subsubsection{Launching, acceleration, collimation}
The three regions in Fig.\ \ref{regions} play different roles in the formation of the jet. At the surface of the disk, a transition  takes place from the high-$\beta$ interior to the magnetically dominated atmosphere of the disk. This is also the region which determines the amount of mass flowing into the jet: it is the {\em launching} region (Sect. \ref{launch}). At some height in the atmosphere the flow reaches the sound speed (more accurately: the slow magnetosonic cusp speed $v_{\rm c}$ given by $v_{\rm c}^2=\cs^2\va^2/(\cs^2+\va^2)$, cf. \cite{heinemann}).  If the gas density at this point is $\rho_0$, the mass flow rate is $\dot m\approx \cs \rho_0$, just as in standard stellar wind theory (cf. \cite{mestel,spruit1}). 

When the temperature in this region is high, for example due to the presence of a hot corona, the atmosphere extends higher above the disk surface and it is easier to get a mass flow started. If the disk atmosphere is cool (temperature much less than the virial temperature), the gas density declines rapidly with height and the mass flow rate becomes a sensitive function of physical conditions near the disk surface. Since these cannot yet be calculated in sufficient detail for realistic disks, the mass flow rate is usually treated as an external parameter of the problem. This is discussed further in Sect. \ref{launch}. 

After launching, the flow is first accelerated by the centrifugal effect, up to a distance of the order of the Alfv\'en radius. The flow velocity increases approximately linearly with distance from the rotation axis (Fig.\ \ref{propts}). 

For acceleration by the centrifugal mechanism to be effective, the field lines have to be inclined outward: the centrifugal effect does not work on field lines parallel to the rotation axis. In the magnetic acceleration model, the high degree of collimation\footnote{Collimation is meant here in the same sense as in optics: the angle measuring the degree to which the flow lines in the jet are parallel to each other. This is different from the {\em width} of the jet (a length scale).} observed in some of the most spectacular jets  must be due to an additional process beyond the Alfv\'en surface (Sect. \ref{colli}).
It is conceivable that this does not happen in all cases: less collimated flows may also exist. They would be harder to detect, but have already been invoked for observations such as the `equatorial outflows' in SS 433 \cite{spencer}, and inferred from the rapidly varying optical emission in the accreting black holes GX 339-4 and KV UMa \cite{kanbach}.

The transfer of energy powering the outflow is thus from gravitational energy to kinetic energy of
rotation, and from there to kinetic energy of outflow via the magnetic field. Note that in the
centrifugal picture the magnetic field plays an energetically passive role: it serves as a conduit
for energy of rotation, but does not itself act as a source of energy. The function of the
magnetic field in the acceleration process can also be viewed in a number of different ways, however; 
this is discussed further below.

\begin{figure}
\centering
\vspace{0.7\baselineskip}
\includegraphics[width=1.03\hsize]{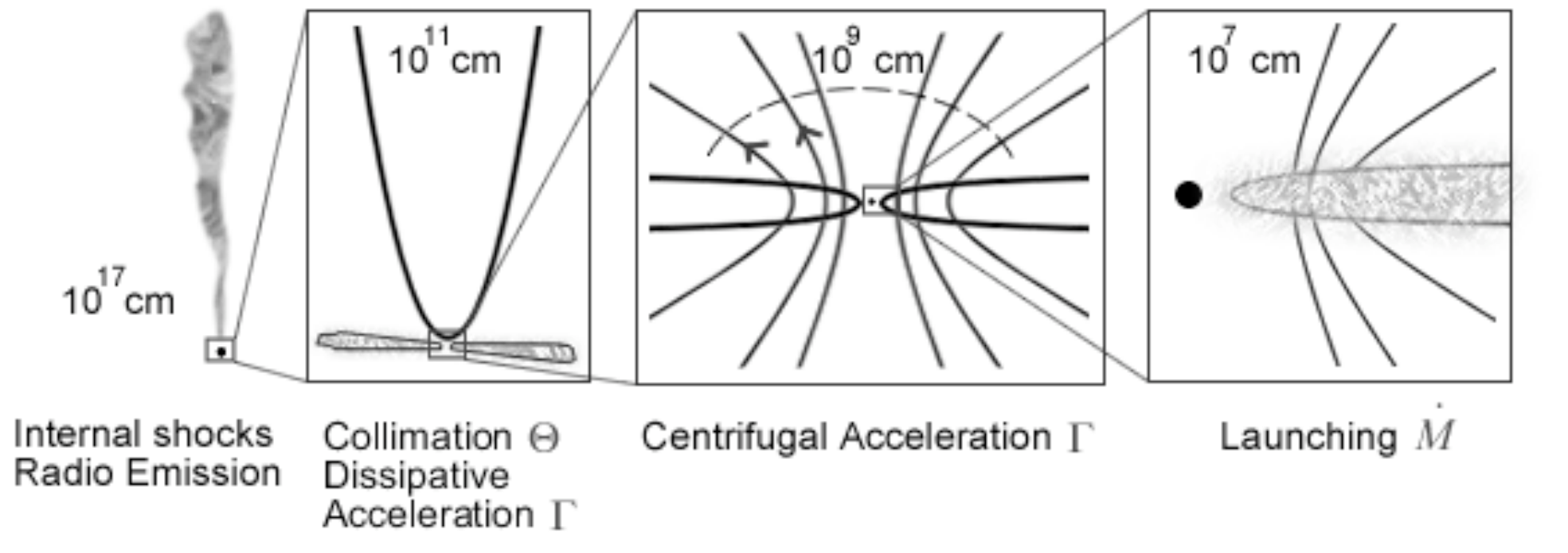}
\vspace{0.3\baselineskip}
\caption{Length scales in a microquasar jet (very schematic). Processes determining the mass flow $\dot M$ in the jet take place in the `launching region' close to the black hole ($\sim 10^7$ cm), but the processes determining the final Lorentz factor ($\Gamma$) and opening angle ($\theta$) may take place on much larger scales.}
\label{lengths}     
\end{figure}

\section{Length scales}
\label{len}

The energy release powering a relativistic outflow happens near the black hole, say $10^7$ cm in the
case of a microquasar. The narrow jets of microquasars seen at radio-wavelengths appear on scales of
the order $10^{17}$ cm. In other words, on scales some ten orders of magnitude larger. It is quite
possible that some of the jet properties are determined on length scales intermediate between these
extremes, at least in some cases. In \cite{spruit3}, for example, we have argued that
collimation of the flow may actually take place on scales large compared with the Alfv\'en radius,
at least in very narrow jets. In Sect. \ref{dissip}, it is shown that such intermediate length scales can also be crucial for acceleration to high Lorentz factors, besides the region around the Alfv\'en radius that plays the main role in the axisymmetric centrifugal acceleration process. 

Much of the current thinking about the processes of launching, acceleration and collimation of the
jet is based on previous analytical models. Numerical simulations of magnetic jets are now becoming
increasingly realistic and useful. They are, however, quite restricted in the range of length scales
and time scales they can cover. This leads to a bias in the interpretation of such simulations: the
tendency is to assume that all steps relevant to the final jet properties happen within the
computational box (cf. Sect. \ref{diffic}). This bias is likely to persist as long as simulations covering realistic ranges in length and time scale are impossible. 

\section{Magnetic jets }

\subsection{Power sources, composition of the jet}
Jets powered by the rotation of a black hole (Blandford-Znajek mechanism) are often assumed to consist of electron-positron pair plasmas, while outflows from rotating disks are regarded as consisting of a normal ion-electron plasma. These associations are not exclusive, however. Since isolated black holes cannot  hold a magnetic field, a field threading the hole requires the presence of an accretion disk holding it in place. Hence it is quite likely that (part of) the jet accelerated by the hole is actually fed with mass from the disk, rather than a pair plasma generated in situ. The simulations by \cite{villiers,mckinney1} are examples of this.

The opposite may also happen. A strong field threading a thin (cool) disk will not be easily loaded with mass from the disk unless the field lines are sufficiently inclined outward, away from the vertical \cite{blandfordpayne}. In addition, the mass loading decreases with increasing field strength, for a given field line geometry \cite{ogilvie}. If too little mass is loaded onto the the field lines, the MHD approximation may not hold. The field lines rotating in a (near) vacuum may then produce a pair plasma above the disk, like the relativistic pair plasma outflows from pulsars. This case has not received much attention so far, perhaps because it would be as difficult to calculate as pulsar winds.

In the literature, the phrase `Poynting flux' is sometimes associated specifically with relativistic
and/or pair-dominated flows. It applies quite generally, however: equally to relativistic and
nonrelativistic flows, and independent of their composition. See also Sect. \ref{accel} below. 

\subsection{`Centrifugal' vs. `magnetic' vs. `Poynting flux'}

\label{accel}
The physical description of the flow-acceleration process has been a source of confusion. There are alternatives to the centrifugal picture sketched above: descriptions in terms of magnetic forces or in terms of `Poynting flux conversion'. The different descriptions are largely equivalent, however; which one to take is a matter of personal taste, or the particular aspect of the problem to be highlighted. A term like  `Poynting jet' for example, does not refer to a separate mechanism, but rather to a particular point of view of the process.

In a frame of reference corotating with the anchoring point of a field line, the flow is everywhere
parallel to the magnetic field (e.g. \cite{mestel}). The component of the Lorentz force parallel to 
the flow vanishes in this frame. There is no magnetic force accelerating the flow: the role of the 
Lorentz force is taken over by the centrifugal force. This is sometimes viewed as a contradiction 
for a magnetic model of acceleration: can one still call the acceleration magnetic if there is no 
work done by magnetic forces?

If the same process is evaluated in an inertial frame, the centrifugal force is absent. Instead, one 
finds that in this frame the flow is accelerated by a force associated with the azimuthal
component of the magnetic field: ${\bf F}=-\nabla B_\phi^2/8\pi -{\bf e}_\varpi B_\phi^2/(4\pi\varpi)$, where $\varpi$ is the distance from the axis\footnote{The poloidal field component  ${\bf B}_{\rm p}$ is absent from the accelerating force. Explanations invoking acceleration of the flow by ${\bf B}_{\rm p}$ are erroneous since the poloidal velocity is parallel to ${\bf B}_{\rm p}$ in a steady flow.}. 

The two descriptions, magnetic and centrifugal, are thus
mathematically equivalent: they are related by a simple frame transformation. The centrifugal 
picture is an elegant way to visualize the acceleration as long as the magnetic field lines corotate with their anchoring point. Where they do not corotate, the field gets wound up into a predominantly azimuthal field, the centrifugal picture looses its meaning, and the acceleration is described 
most simply in terms of the forces exerted by the azimuthal field component $B_\phi$.

Finally, the same process can also be viewed as the conversion of a Poynting flux of electromagnetic
energy into kinetic energy. To see this, recall that in magnetohydrodynamics the electric field $\bf
E$ is given by $E={\bf v\times B}/c$, so that the Poynting flux
\beq {\bf S}={c\over 4\pi}{\bf E\times B}\eeq
can be written as
\beq {\bf S}={\bf v}_{\perp}B^2/(4\pi),\label{mhdpoy}\eeq
where ${\bf v}_{\perp}$ is the component of the flow velocity perpendicular to the magnetic field.  This shows that it is not necessary to think of Poynting flux as EM waves in vacuum. It applies equally well in MHD, and not only to waves but also to stationary magnetic flows. 

Expression (\ref{mhdpoy}) can be interpreted as a flux of magnetic energy, advected with the fluid, in a direction perpendicular to the field lines\footnote{The actual flux of magnetic energy would of course be ${\bf v}_{\perp}B^2/8\pi$. The `missing' $B^2/8\pi$ represents the `$PdV$-work' done by the source of the flow against the magnetic pressure at the base of the flow. See also Sect. \ref{dissip}}. Borrowing a useful analogy from hydrodynamic flows, the Poynting flux in MHD plays the role of a `magnetic
enthalpy flux'. The centrifugal acceleration process is equivalent to the gradual (and incomplete) 
conversion of Poynting flux into a flux of kinetic energy, much like the
conversion of enthalpy into kinetic energy in an expanding hydrodynamic flow. Near the base of the
flow (for example the surface of the accretion disk which supplies the mass flux into the wind), the
enthalpy flows almost entirely in the form of a Poynting flux $\bf S$. S declines gradually with
distance and the kinetic energy increases correspondingly, most of the energy transfer taking place
around the Alfv\'en radius (in the axisymmetric case, see however Sect. \ref{dissip}). 

\subsection{Poynting flux conversion efficiency: axisymmetric}
\label{coneff}
Since the flow is magnetic everywhere, at least some of the energy is carried in the form of a
magnetic energy flux. The work done by the central engine is not converted completely into
kinetic energy, and one may wonder what determines the efficiency of conversion of Poynting flux. 

It turns out that the answer depends critically on the symmetry conditions imposed on the flow. When 3-D, nonaxisymmetric processes are allowed, conversion can be more efficient than in axisymmetric
flows. This is discussed further in Sect. \ref{dissip}. Since much of the current views are still
based on axisymmetric models, however, consider these first. 

If $S_0$ is the Poynting flux at the base of the flow ($\sim$ the power output of the jet) and $F_{\rm K}$ the kinetic energy flux, we can define this efficiency $f$ as
\beq f=F_{{\rm K}\infty}/S_0=F_{{\rm K}\infty}/(F_{{\rm K}\infty}+S_\infty),\eeq
where $_\infty$ denotes the asymptotic values at large distance. A simple
model for which this can be calculated is the cold Weber-Davis model (for introductions see \cite{mestel,spruit1}; for a concise and elegant mathematical treatment see \cite{sakurai85}). In this model the poloidal field lines are straight and radial, so that the poloidal field strength varies as $1/r^2$ (a `split monopole'). In this approximation the azimuthal field and the flow can be calculated exactly, but  the force balance in the latitudinal ($\theta$) direction is neglected.  In the `cold' version of the model, the gas pressure is also neglected. In the nonrelativistic limit, the conversion efficiency in this model is of order unity, so a significant fraction of the power delivered by the central engine remains in the flow as magnetic energy. 

In the relativistic case, i.e. when the flow reaches large Lorentz factors ($\Gamma$), a
smaller fraction of the Poynting flux is converted to kinetic energy than in the non-relativistic case. This can be illustrated with the relativistic extension of the Weber-Davis model, given already by Michel's model\cite{michel,goldreich}.   This model gives an exact solution for a relativistic, magnetically accelerated flow, in the approximation of a purely radial geometry  for the poloidal field. 

If ($r,\theta,\phi$) are spherical coordinates centered on the source of the flow, the simplest case to visualize is a flow near the equatorial plane, $\theta=\pi/2$. The split monopole assumption for the poloidal field components implies $v_\theta=B_\theta=0$. As in the nonrelativistic case, the field at large distances is nearly exactly azimuthal. The radial component of the Lorentz force is then 
\beq F_r=-\partial_r B^2_\phi/(8\pi)-B^2_\phi/(4\pi r).\label{balan}\eeq
From the induction equation one finds that 
\beq B_\phi r v_r={\rm cst,}\label{fiflux}\eeq
i.e. the `flow of azimuthal field lines' is constant. Asymptotically for $\Gamma\rightarrow\infty$,
$v\approx c$, so $B_\phi\sim 1/r$. The two terms in the Lorentz force then cancel. Equation \ref{balan}
holds at all latitudes, in this split monopole configuration. 

The consequence of this cancellation is that a flow in a purely radial poloidal field stops being accelerated as soon as it develops
a significant Lorentz factor. From then on acceleration and conversion of Poynting flux slow down.
Moderately efficient conversion of Poynting flux to kinetic energy is possible, but only up to
modest Lorentz factors. High Lorentz factors are also possible, but at the price of converting only
a small fraction of the energy flux. This is seen in the expression for the terminal Lorentz factor
$\Gamma_\infty$ in Michel's model:
\beq \Gamma_\infty\approx {m}^{1/3},\eeq
where $m$ is Michel's magnetization parameter,
\beq m=B_0^2/(4\pi\rho_0 c^2),\eeq
and $B_0$, $\rho_0$ are the magnetic field strength and mass density at the base of the flow (where
it is still non-relativistic). If conversion of Poynting flux into kinetic energy were complete, it
would produce a flow with Lorentz factor $\Gamma_{\rm c}$,
\beq \Gamma_c=m.\eeq
The actual efficiency of conversion is thus
\beq f\equiv \Gamma_\infty/\Gamma_{\rm c}\approx m^{-2/3}\approx 1/\Gamma_\infty^2.\eeq
This is a small number if large Lorentz factors are to be achieved. One gets either good conversion of
Poynting flux into kinetic energy, or large terminal speeds but not both.

\subsubsection{Conversion in  diverging flows}
\label{diverg}
The conclusion from the previous subsection holds under the `split monopole' assumption that (apart from its azimuthal component) the flow expands exactly radially. If this is not the case, the cancellation is not exact, and continued acceleration possible. To achieve this, the magnetic
pressure gradient term in (\ref{balan}) has to be larger, relative to the second term, than it is in
the split monopole geometry. This is the case when the azimuthal field {\em decreases
more rapidly} with distance, in some region of the flow (c.f. \cite{phinney,begelman}). This is perhaps the 
opposite of what intuition would tell, but similar to the acceleration of supersonic flows in expanding nozzles. 

To make this a bit more precise, consider the magnetic forces in a steady axisymmetric flow. The flow can be considered separately along each poloidal field line. Let $z$ be the distance along the axis of the jet, and $R(z)$ the cylindrical radial coordinate of a field line. Let $d(z)$ be the distance, in a meridional plane, between two neighboring poloidal field lines. Consider distances far enough from the Alfv\'en radius that the rotation rate of the flow has become negligible. The field has then become purely azimuthal. If the flow is steady, it follows from the induction equation that
\beq vB_\phi d=~cst. \label{bphiflo}\eeq
i.e., the `flow rate of azimuthal field loops' is constant along a flow line. To investigate under which conditions the flow is accelerated, consider the component $F_s$ of the Lorentz force along the flow line. It is the sum of a curvature force and a magnetic pressure gradient. Instead of calling the equation of motion into action, it is sufficient to evaluate the forces under the assumption that the flow speed is constant. Acceleration is then indicated if there is a net outward force under this assumption.

If $\theta$ is the angle ${\mathrm{atan}(d}R/{\rm d} z)$ of the field line with the axis, the component of the curvature force along the flow line is $-\sin\theta B_\phi^2/(4\pi R)$ and the magnetic pressure gradient along the flow is $-\cos\theta{\rm d}B_\phi^2/{\rm d} z/8\pi$. Summing up while using (\ref{bphiflo}) yields
\beq F_s={B_\phi^2\over 4\pi}{{\rm d}\over{\rm d}s}\ln(d/R), \eeq
where ${\rm d/d} s=\cos\theta{\rm d/d}z$ is the derivative along the flow line.
For a purely radial flow, $d\sim R$, and the force vanishes as expected. The result is more general, however: it applies to any `homologous' flow, with $d/R$ independent of $z$. That is, it holds if the distance $d$ between poloidal field lines varies in the same way as the distance $R$ of the field line from the axis.  

For net acceleration it is thus not sufficient that the overall expansion of the flow is more rapid than radial. Expansion must also be non-selfsimilar: acceleration takes place only on field lines that diverge from their neighbors faster than the overall expansion of the flow. This is unlike the case of the `nozzle effect' in supersonic hydrodynamical flows. 

The result can also be derived by considering the Poynting flux $\bf S$ itself. In ideal MHD it is given by Eq. (\ref{mhdpoy}): ${\bf S}={\bf v}_\perp B^2/4\pi$   where ${\bf v}_\perp$ is the velocity component perpendicular to the flow. Again assuming that the field is already purely azimuthal, $S$ is parallel to the flow as well, ${\bf S=v}B_\phi^2/4\pi$.
The total Poynting flux flowing between two poloidal surfaces at distance $R(z)$ from the axis and separated by a distance $d$ as before is then 
\beq S_{\rm p}=S\,2\pi Rd.\eeq
Poynting flux is converted into kinetic energy if  this quantity decreases with distance along the flow. Again using (\ref{bphiflo}) to eliminate $B_\phi$:
\beq S_{\rm p}={k^2\over 2v}{R\over d},\eeq
where $k=vB_\phi d$, the flow rate of the azimuthal field. It follows that for Poynting flux conversion it is necessary for $d$ to increase with distance $z$ faster than $R$, in agreement with the derivation above.

The conclusion is that at distances where the field has already become azimuthal, efficient Poynting flux conversion is not possible by the simplest expanding-nozzle effect. The expansion must be non-homologous, and acceleration takes place only in parts of the flow where the separation between neighboring flow lines increases faster than average. See \cite{moll09,tchek} for further discussion.

The assumption of a purely azimuthal field does not hold closer to the source and up to several times the  Alfv\'en distance, and the above reasoning does not apply there. The region around the fast mode critical point of the flow appears to be conducive to Poynting fux conversion even in homologously expanding flows ($d\sim R$). A detailed study has been given by \cite{daigne}, for the general relativistic case and including the contributions from thermal pressure. Reasonable conversion efficiencies, of order 50\%, are achieved if the opening angle of the flow increases by a factor of a few in the region around the fast mode point (see also the numerical simulation by \cite{barkov}). 

The limitations of this process of acceleration by divergence of the opening angle become more severe when actual physical conditions leading to divergence are considered. A limiting factor is causality. In a flow of Lorentz factor $\Gamma$, parts of the fluid moving at angles differing by more than $\theta_{\rm max}=1/\Gamma$ are causally disconnected: there is no physical mechanism that can exchange information between them. Hence there are no physically realizable processes that can cause them to either converge to, or diverge from each other. Jets accelerated by flow divergence therefore must satisfy 
\beq \theta\Gamma <1,\label{begam}\eeq 
where $\theta$ is the asymptotic opening angle of the jet. Jets in Gamma-ray bursts, with inferred opening angles of a few degrees and minimum Lorentz factors of $\approx 100$, can thus not be accelerated by homologous expansion. 

Condition (\ref{begam}) is less constraining in the case of AGN jets, with inferred Lorentz factors in the range 3-30. Since efficient Poynting flux conversion by this process requires an expansion in opening angle by a factor of several, however, the initial opening angle of an AGN jet would have to be several times smaller than the observed angles. It is not clear if this is compatible with observations.
In Sect. \ref{dissip} an efficient conversion process is presented that does not require divergence of the flow  (in fact it works best in a converging flow geometry), and is applicable both to AGN and GRB.

\subsubsection{Conversion efficiency: possible artefacts}
Because of the near cancellation of terms in the magnetic acceleration, some caution is needed when
interpreting results of magnetically driven flow calculations, whether they are analytic or
numerical. In analytic models,  simplifying assumptions can tip the balance in
favor of one of the two terms, leading to a spurious acceleration or deceleration. 

In numerical work, the unavoidable presence of numerical diffusion of field lines (e.g. lower order 
discretization schemes) can cause artificial acceleration. If such diffusion is effective, i.e. the numerical resolution low, the toroidal field can potentially decay by annihilation across the axis. This would result in a  decrease of the magnetic pressure with distance along the axis, increasing the pressure gradient term in (\ref{balan}), also resulting in acceleration. The signature of such an artefact would be that the acceleration found decreases as numerical resolution is improved. 

Acceleration by a decrease in magnetic energy  along the flow is itself a real effect, however, if a mechanism for dissipating magnetic energy within the jet is present. I return to this in Sect. (\ref{dissip}).

\subsection{`Magnetic towers'}
Another picture of magnetic jets produced by a rotating, axisymmetric source is that of a `magnetic
tower', sometimes presented as an intrinsically separate mechanism with its own desirable properties. It is a simplified picture of the magnetic acceleration process, in which the magnetic field is depicted as a cylindrical column of wound-up magnetic field. One loop of toroidal field is added to the column for each rotation of the footpoints (cf. Fig.\ \ref{windup}). The tower is assumed to be in pressure balance with an external confining medium. The attraction of this model is that it is easily visualizable. In addition, some of the numerical simulations look much like this picture. This the case in particular for  simulations done in a cylindrical numerical grid. 

The description of a magnetic jet as a `magnetic tower' does not address the acceleration of the flow, nor does it address how a jet is collimated. It is a kinematic model describing the shape of the field lines once a flow has been assumed. To explain acceleration of the flow, the dynamics of the centrifugal and dissipative acceleration mechanisms described above and below have to be included.

\subsection{Flows with high mass flux}
\label{hiflo}
At low mass flux in the wind, the Alfv\'en surface is at a large distance from the disk\footnote{Except in relativistic disks: in this case the Alfv\'en surface approaches the light surface (`cylinder') corresponding to the rotation rates of the footpoints on the disk.}, and it moves inward with increasing mass flux. One may wonder what happens when so much mass is loaded onto the field lines that the field is too weak to enforce corotation. When conditions in the wind-launching zone (Sect. \ref{launch}) produce such a high mass flux, the `centrifugal' acceleration picture does not apply any more. Cases like this are likely to be encountered in numerical simulations, since the opposite case of low mass flux is much harder to handle numerically. Low mass fluxes cause problems associated with the high Alfv\'en speeds in the accelerating region and the larger computational domain needed, so the characteristic behavior of a centrifugal wind, with Alfv\'en surface far from the source, is not the first one expects to encounter in simulations.

The high mass flux case can be illustrated with an analytical model: the `cold Weber-Davis' model mentioned above. Consider for this a flow with the poloidal component of the $B$-field purely radial ($B_\theta=0$) near the equatorial plane of a rotating source (i.e. the plane of the disk). Define the mass-loading parameter $\mu$ as
\beq \mu=\rho_0{4\pi v_0\Omega r_0\over B^2_0}={v_0\Omega r_0\over v^2_{{\rm A}0}},\eeq
where  $\Omega$ is the rotation rate of the field line with footpoint at distance $r_0$ from the axis, $v_0$ and $v_{{\rm A}0}$ the flow velocity and Alfv\'en speed at $r_0$. The solution of the model then yields for the Alfv\'en distance $r_{\rm A}$:
\beq r_{\rm A}=r_0[{3\over 2}(1+\mu^{-2/3})]^{1/2}. \eeq
Various other properties of the flow can be derived (see the summary in Sect. 7 of Spruit 1996). An example is the asymptotic flow speed:
\beq {v_\infty\over \Omega r_0}=\mu^{-1/3}. \label{hsvinf}\eeq

\begin{figure}
\centering
\includegraphics[width=0.8\hsize]{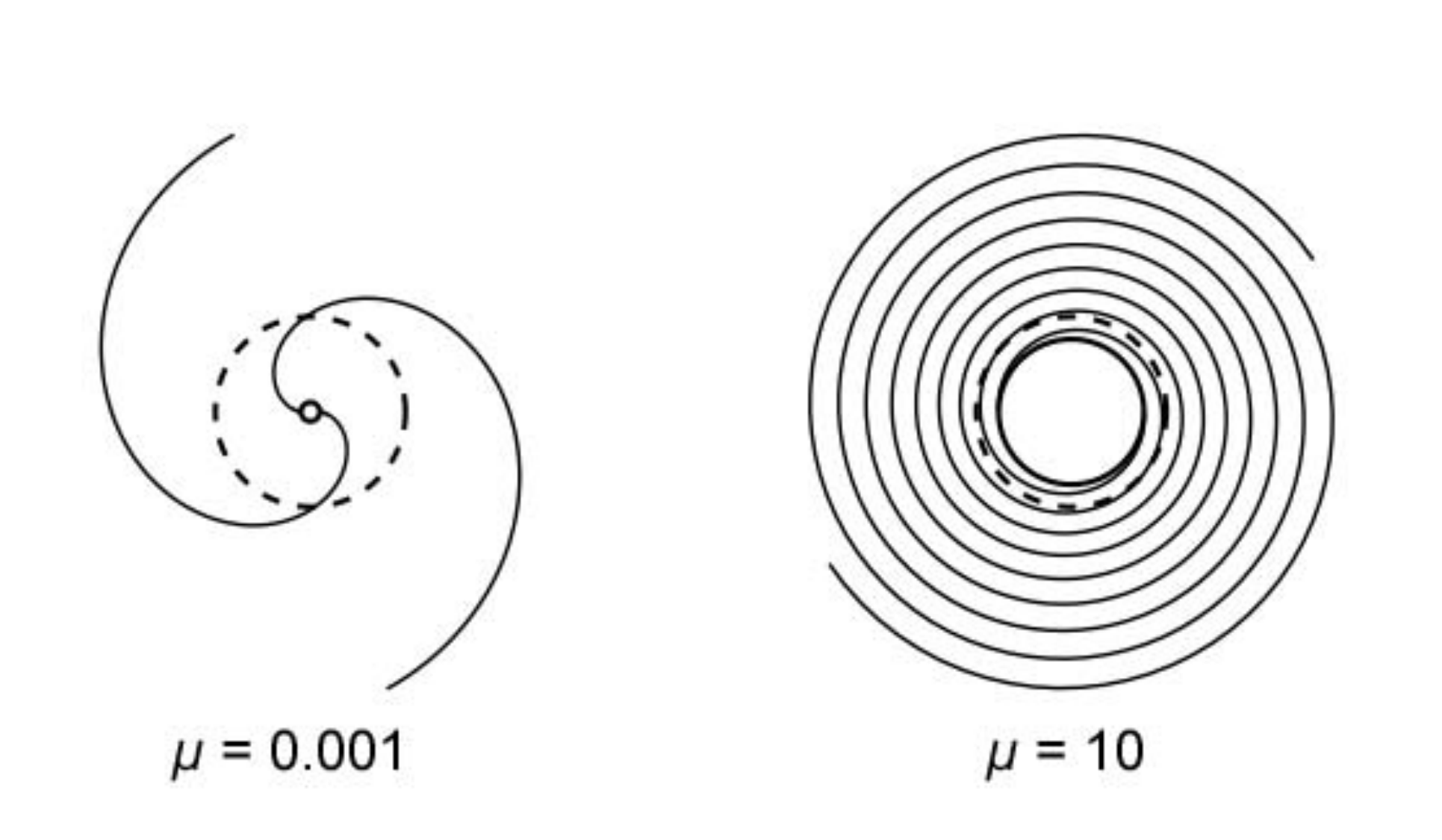}
\caption{Shape of the magnetic field lines of the cold Weber-Davis wind model, for two values of the mass-loading parameter $\mu$. At low mass loading, the Alfv\'en radius $r_{\rm A}$ (dashed) is far from the source surface (solid circle), the angle of the field lines at $r_{\rm A}$ is of order unity. At high mass flux, the field lines are already wound up into a tight spiral before the flow reaches $r_{\rm A}$. The flow is slow in this case, and it is likely to be subject to various nonaxisymmetric instabilities.}
\label{coldwd}     
\end{figure}

Figure \ref{coldwd} shows how the field lines are wound up for a low mass flux and a high mass flux case. These scalings have been derived only for the rather restrictive assumptions of the cold Weber-Davis model. It turns out, however,  that they  actually hold more generally, at least qualitatively; they have been reproduced in 2-D numerical simulations \cite{anderson}. 

Equation (\ref{hsvinf}) shows that at high mass flux, the outflow speed goes down, as expected, and for $\mu>1$ actually drops below the orbital velocity at the footpoint. At such low velocities, the travel time of the flow, say to the Alfv\'en radius, becomes longer than the crossing time $t_{\rm A}$ of an Alfv\'en wave, $t_{\rm A}=r/v_{\rm A}$. There is then enough time for Alfv\'enic instabilities to develop, such as buoyant (Parker-) instability and/or kink modes before the flow reaches substantial speed, interfering with the acceleration process. Since these are nonaxisymmetric, such effects do not show up in the typical 2-D calculations done so far. It is quite possible that the high mass-flux case will turn out to be highly time-dependent, and poorly represented by the above scalings derived for steady flow. 3-D simulations of high mass flow cases would therefore be interesting, but likely to be challenging.

High mass flux conditions may well occur in astrophysical objects, but they probably will not produce the familiar highly collimated high-speed jets. They might be involved in slower `equatorial' outflows inferred in some objects.

\section{Ordered magnetic fields}
\subsubsection{Impossibility of generation by local processes}

An {\em ordered} magnetic field such as sketched in Fig.\ \ref{regions} is usually assumed in work on magnetocentrifugal acceleration: a field of uniform polarity threading the (inner regions of) a disk. This is sometimes chosen as a
representative idealization of more complicated configurations such as might result from magnetic
fields generated in the disk. The ordered configuration has the advantage of simplicity: all
field lines anchored in the disk extend to infinity, and the flow can be a smooth function of
distance from the axis. If the field is not of uniform polarity, some field lines form closed loops
connecting parts of the disk surface instead of extending to infinity, and the outflow will be patchy (cf. Fig.\ 5 in \cite{blandfordpayne}). 

A field of uniform polarity, however, is subject to an important constraint: it cannot be created in
situ by local processes in the disk. It can only exist as a consequence of either the initial
conditions, or of magnetic flux entering or leaving the disk through its outer (radial) boundary. To
see this formally, consider a circle at $r=R, z=0$ (in cylindrical coordinates $r,\phi,z$
centered on the disk), where $R$ could be the outer radius of the disk, or the radial outer boundary
of the computational domain. Let $S$ be a surface with this circle as its boundary, with normal
vector ${\bf n}$, and let $\Phi=\int{\rm d }S\, {\bf B\cdot n}$ the magnetic flux through this
surface. On account of ${\rm div}\,{\bf B}=0$, $\Phi$ is independent of the choice of $S$, as long as
the boundary at $r=R$ is fixed, and we can take $S$ to be in the midplane $z=0$ of the disk. With
the induction equation, we have
\beq \partial_t\Phi=\int{\rm d}r{\rm d}\phi~ r[{\bf\nabla\times(u\times B)}]_z.\eeq
With $u_r(0,\phi,z)=B_r(0,\phi,z)=0$ by symmetry of the coordinate system, this yields
\beq \partial_t\Phi=-\int{\rm d}\phi~ R[u_zB_r-u_rB_z],\label{flux}\eeq
where the integrand is evaluated at $r=R$. The square bracket can be written as $u_\perp B_{\rm p}$,
where ${\bf B}_{\rm p}$ is the poloidal field $(B_r,B_z)$ and ${\bf u}_\perp$ the velocity component
perpendicular to it. The RHS of (\ref{flux}) can thus be identified as the net advection of poloidal
field lines across the outer boundary. 

If this flow of field lines across the outer boundary vanishes, the net magnetic flux $\Phi$ through
the disk is constant. It vanishes if it vanishes at $t=0$: it can not be created by local processes in the disk, including large scale dynamos (even if these were to exist in accretion disks). 

The magnetic flux through a disk a is therefore {\em global} quantity rather than a local function of conditions near the center. It depends, if not on initial conditions, on the way in which magnetic flux is transported through the disk as a whole. Since it is not just a function of local conditions in the disk, it acts as a {\em second parameter} in addition to the main global parameter, the accretion rate. This has an interesting observational connection: not all disks produce jets, and the ones that do, don't do it all the time. The possibility suggests itself that this variation is related to variations in the magnetic flux parameter of the disk \cite{spruit5}. 

\subsubsection{Field strengths}
\label{strength}
A particular attraction of ordered fields is that they can be significantly stronger than the fields produced by magnetorotational (MRI) turbulence. The energy density in MRI fields is limited to some (smallish) fraction of the gas pressure  at the midplane of the disk. The exact fraction achievable still appears to depend on details such as the numerical resolution, with optimistic values of order 0.1 commonly quoted, while values as low as 0.001 are being reported from some of the highest resolution simulations \cite{fromang}.  

The strength of ordered fields  can be significantly higher, limited in principle only by equipartitition of magnetic energy with orbital kinetic energy, or equivalently by the balance of magnetic forces with gravity. In practice, interchange instabilities already set in when the fractional support against gravity reaches a few per cent, as shown by the numerical simulations of \cite{stehle}. For a thin disk, however, this can still be substantially larger than equipartition with gas pressure,  since the orbital kinetic energy density is a factor $(r/H)^2$ larger than the gas pressure at the midplane. Magnetic fields of this strength actually suppress magnetorotational instability. Instead, their strength is limited by new instabilities driven magnetic energy rather than the shear in the orbital motion (see discussion and results in \cite{stehle}). 

Strong fields are also indicated by the observations of rapidly varying optical emission in some accreting black holes, in particular GX 339-4 and KV UMa. As argued in \cite{fabian}, the only realistic interpretation of this radiation is thermal synchrotron emission from a compact region near the black hole. The inferred optical depth requires very strong magnetic fields \cite{kanbach},
probably larger than can be provided by MRI turbulence.

\subsection{Ordered magnetic fields in numerical simulations}
Equation (\ref{flux}) says that the net magnetic flux $\Phi$ through the disk does not change unless
there is a net advection of field lines into or out of the disk boundary. This implies that the
velocity field inside the disk cannot create a net poloidal flux, no matter how complex or
carefully construed the velocities. 

A bundle of ordered magnetic flux threading a black hole, such as seen in simulations (e.g. \cite{villiers,mckinney1,hawley}), can only have appeared `in situ' by violation of $\rm {div}\,{\bf B}=0$. Since this is unlikely with the codes used, the flux bundles seen must have developed from flux that was already present at the start of the simulation.  

\begin{figure}
\centering
\includegraphics[width=\hsize]{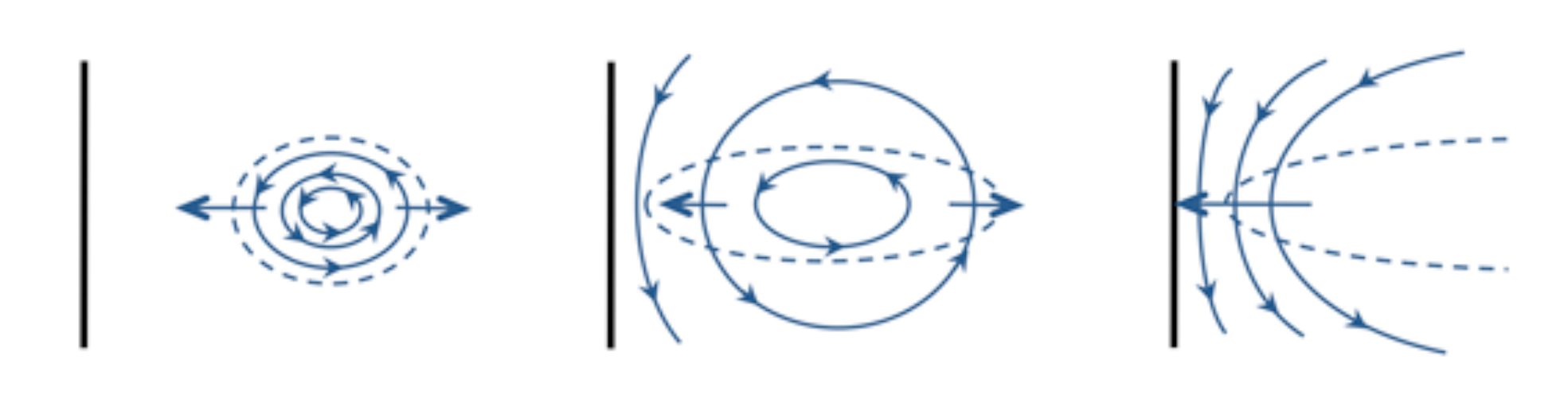}
\caption{Formation of a central flux bundle by a spreading torus containing loops of poloidal field}
\label{spread}     
\end{figure}

The way this happens has been pointed out in \cite{igu} and \cite{spruit6}; it can be illustrated with the simulations of  \cite{mckinney1}. The initial state used there is a torus of mass, with an initially axisymmetric field consisting of closed poloidal loops. The net poloidal flux through the midplane of the calculation thus vanishes: downward flux in the inner half of the torus is compensated by upward flux in the outer part (Fig.\ \ref{spread}). The differential rotation in the torus generates MRI turbulence, causing the torus to spread quasi-viscously. The inner parts spread towards the hole while the outer parts spread outwards. The magnetic loops share this spreading: the downward flux in the inner part spreads onto the hole, the upward flux spreads outward.

This explains the formation of a flux bundle centered on the hole in the simulations, but also makes clear that  the result is a function of the initial conditions. The process as simulated in this way, starting with a torus close to the hole, is not really representative for conditions in an extended long-lived accretion disk. If  a poloidal loop as in the initial conditions of present simulations were present in an actual accretion flow,  the downward part of the flux would accrete onto the hole as well, canceling the flux threading the hole.

The dependence on initial conditions is demonstrated more explicitly by some of the results in \cite{machidab} and \cite{villiers}. These results show that  different initial conditions (a toroidal instead of a poloidal field) result in very similar magnetic turbulence in the disk, but without an ordered Poynting flux jet.  See also the recent discussion in \cite{beckwith}.

This disagrees with  earlier  suggestions  \cite{villiers,krolik} that such jets would be a natural generic result of MRI-generated magnetic fields, or even the claim  \cite{mckinney3} that  a net magnetic flux through the disk would appear from MRI turbulence. 

Existing simulations thus show that a flux bundle at the center of a disk can form from appropriate
initial conditions, but leave open the question which physics would lead to such conditions. As I
argue below, this need not be seen only as an inconvenience. The same question may well be related to the puzzling phenomenology of X-ray states in X-ray binaries. At the same time, the simulation results are important since they appear to demonstrate that there is a flaw in the analytical models previously used for the accretion of net magnetic flux through a disk. This is discussed further in the next Section.

\subsection{Accretion of ordered magnetic fields?}
\label{accord}
If magnetic fields of net polarity cannot be created internally in a disk, but a net polarity at
the center of the disk is still considered desirable, there are two possibities:\\
- The field is accreted from the outside (a companion, or the interstellar medium),\\
- It forms by systematic separation of polarities somewhere in the disk.

The first of these has been a subject of several studies, the conclusions of which are
discouraging. The model used is that of a diffusing disk, where angular momentum transport
is mediated by a viscosity $\nu$, and the magnetic field diffuses with diffusivity $\eta$. If both result from some quasi-isotropic turbulence, they are expected to be of similar magnitude. Numerical simulations addressing this question \cite{guan} show that the ratio $\nu/\eta$ (the magnetic Prandtl number) is close to unity.

Vertical field lines are accreted through the disk at a rate $\sim\nu/r$, while diffusing outward at a rate $\sim\eta/r$. In a steady state the balance between the two would yield a strong increase of field strength towards the center of the disk. The assumption of a vertical field being accreted is very unrealistic, however, since accreted field lines cannot stay vertical. In the vacuum above the disk the field lines bend away from the regions of strong field, so that the field lines make a sharp bend on passing through the disk. As shown first by \cite{vanballegooijen}, the result is that magnetic flux is accreted very inefficiently. Because of the sharp bend, the length scale relevant for magnetic diffusion is the disk {\em thickness} $H$, rather than $r$, and diffusion correspondingly faster. The result is that accretion in a disk with $\nu\approx\eta$ cannot bend field lines by more than an angle $\sim H/r$ \cite{lubow1}, and the increase of field strength toward the disk center is negligible. 

This result would seem to exclude the accretion of magnetic flux to the center of a disk by amounts significant enough to create a strong ordered field around the central mass. The observational indications for the existence of such fields are nevertheless rather compelling. An attempt to circumvent the diffusion argument above was made by \cite{spruit5}. We appealed there to the fact that in addition to the external field to be accreted, the disk also has its own, magnetorotationally generated small scale magnetic field. This field is likely to be highly inhomogeneous, with patches of strong field separated by regions of low strength, as seen in recent numerical simulations \cite{machidaa,fromang,guan}. An external field accreted by the disk then does not cross the disk uniformly, but through the patches of strong field. Such patches can effectively loose angular momentum through a wind, thereby beating the diffusion argument and causing the external field to be accreted. 

The ability of an accretion flow to maintain a bundle of strong field at its center, first proposed by \cite{kogan74}, is suggested by the simulations of \cite{igu}, \cite{mckinney1} and others, at least for the geometrically thick flows that are accessible with numerical simulations. 

\begin{figure}
\centering
\includegraphics[width=0.6\hsize]{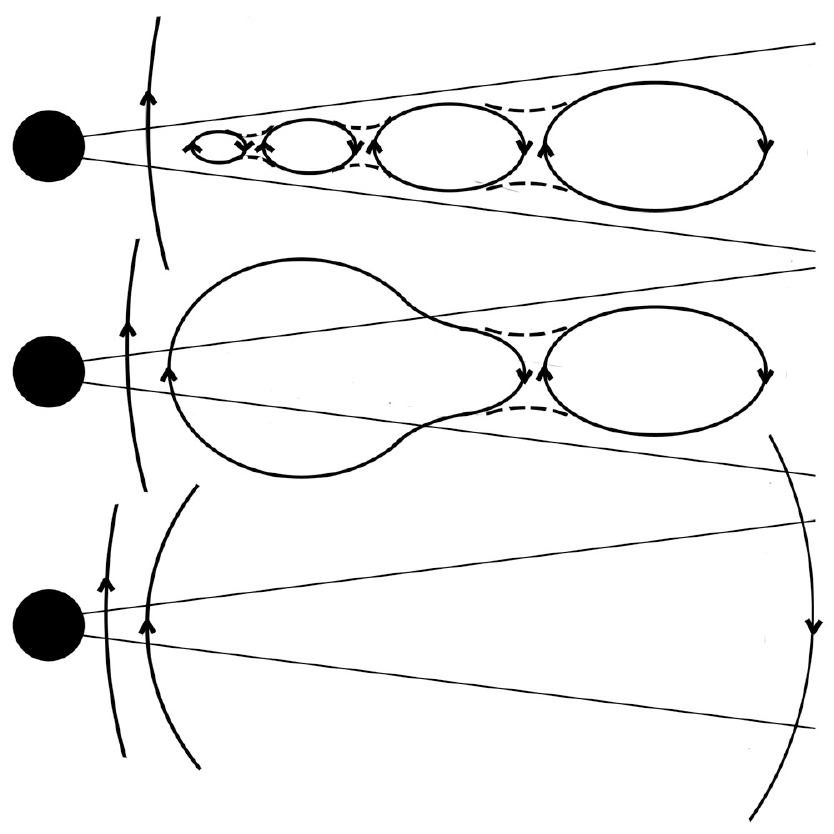}
\caption{Separation of magnetic polarities by coordinated small scale action. If small scale loops created in a disk have the same orientation (arrows), reconnection (dashed lines and middle panel) with subsequent escape from the disk would leave a net flux bundle in the inner disk. }
\label{loopmerge}     
\end{figure}

\subsubsection{Spontaneous separation}
The alternative possibility of a spontaneous separation of magnetic polarities from a mixture (as generated by MRI turbulence, for example) is also possible in principle, if a process of `coordinated small scale action' exists. Assume that some form of magnetic turbulence in the disk contains small scale ($\Delta r\sim H$) loops of poloidal field (in addition to a toroidal field component). If there is a reason why the loops are of the same sign throughout the disk, as sketched in Fig.\ \ref{loopmerge}, reconnection between them can create an ordered radial field. Allowing this field to escape through the upper and lower surfaces of the disk leaves two bundles of field lines of opposite sign crossing the disk, near its inner and outer boundaries. Though this still has not produced a net flux through the disk, it has separated polarities enough that the canceling flux in the outer disk need not influence the flux bundle in the inner disk much. The scenario is unrealistic, however.

\subsubsection{Dynamos}
The assumption of coordinated action on which the scenario  Fig.\ \ref{loopmerge} is based is unrealistic. For the loops to know the orientation needed for the process to work, at a minimum the information communicating this knowledge, the Alfv\'en speed, has to travel through the disk sufficiently fast. The life time of poloidal loops like those in Fig.\ \ref{loopmerge} is of the order of the local orbital period, however. An Alfv\'en wave travels only of the order of a disk thickness in this time. By the time the loop's orientation has been communicated to any larger distance, it has been replaced randomly by another loop. Larger scales thus will be acausal, governed by the statistics of random superposition. Simulations of magnetic turbulence in disks confirm this \cite{guan}.  This calls into question the literature on large scale fields produced with `disk dynamo' equations, which have the assumption of large scale field generation already built into them. The well-established the practice of using  such mean field equations does not replace justification of their applicability.

\subsection{The `second parameter' in accreting X-ray sources}
Previous analytic theory, as well as the recent numerical simulations, show the advantage of an ordered magnetic field near the central object for creating powerful outflows; though less effective forms of outflow associated with random magnetorotationally generated magnetic fields appear possible as well \cite{igu,machidab}.

Ordered fields also make the puzzling behavior of X-ray binaries easier to understand. 
The phenomenology of X-ray binaries (black hole and neutron star systems) includes changes in the X-ray spectrum and the time variability of the X-ray emission (see contributions elsewhere in this volume). For many years the prevailing view in the interpretation of X-ray binaries has been that this phenomenology is governed by a single parameter: the accretion rate (not counting the system parameters of the binary itself). 

This view was supported by the fact that X-ray  binaries with neutron star primaries showed a systematic behavior, with their spectral and timing properties ordered aproximately along a single track in color-color or color-intensity diagrams. It did not agree with the observations of X-ray transients (mostly black hole systems) including the `canonical' black hole transient GU Mus. These do not conform to the movement back and forth along a single track in color-intensity diagrams expected from a single-parameter system, showing instead motion around wide and/or irregular `loops'. 

In spite of this, the phenomenology of these transients has traditionally been interpreted as a single sequence of states with declining accretion rate, the apparent anomalies blamed on, for example, the transient nature of the sources. The X-ray spectrum and the properties of the time variability in these sources shows strong similarities in states of very different brightness (e.g. the `very high' and `intermediate' states, \cite{rutledge,belloni05,HomBel05}). This provides a compelling clue that the phenomenology is not simply a function of the instantaneous accretion rate alone. Anomalies in the neutron star binaries (`parallel tracks' e.g. \cite{vanderklis}), though smaller in magnitude, point in the same direction. 

Instead of just the instantaneous accretion rate, one could imagine that the state of the system depends also on the {\em history} of the accretion rate. This would be the case if there is a physical property of the disk causing hysteresis, such that the state is different during increasing and decreasing accretion rates, for example. Such a mechanism might be an evaporation processes depleting the inner regions of the disk, such that the size of the depleted zone depends on the history of the accretion rate (see review by \cite{homan}). 

A more radical idea is that a true `second parameter' is involved in the state of the accretion flow and its X-ray and timing properties. Apart from binary parameters such as the masses and orbital separation, the mass transfer rate from the companion is the only parameter determining a (steady) hydrodynamic accretion disk. In the influential standard theory of disks based on local viscosity prescriptions, the physical state at a given point of the disk is only a function of the {\em local} mass flux. This makes the theory much more deterministic than the properties of X-ray transients seem to indicate. 

A useful second parameter would therefore preferably be a {\em global} quantity: a property of the disk as a whole, independent of the accretion rate, which varies between disks or with time in a given disk. It is hard to come up with plausible candidates. As argued in \cite{spruit5} a promising one, however, is the {\em net flux} $\Phi$ of field lines crossing the disk. As shown above, this is a truly global parameter: its value is determined only by inheritance from the initial conditions and the boundary conditions at its outer edge; it can not be changed by local processes in the disk. One could imagine, for example, that a large amount of flux would interfere with the accretion process in the inner parts of the disk \cite{kogan74}, so flux could get concentrated there. This could be related  to the nature of the poorly understood hard X-ray state, and the indications for `truncation' of the inner disk (cf. \cite{churazov,dangelo} and references therein). 

Another useful property of the global magnetic flux as a second parameter is the observed relation between X-ray states and the occurence of jets from X-ray binaries. If the hard X-ray state is indeed one with a high magnetic flux in the inner disk, its connection with jets would be natural since current theory  strongly suggests them to be magnetically driven phenomena.

\section{Flow acceleration by magnetic dissipation}
\label{dissip}

In 3 dimensions, the energy carried in the form of the wound-up magnetic field can decay by internal dissipation, something which is excluded in axisymmetry. This turns out to be a very efficient way of accelerating the flow to high Lorentz factors \cite{drenkhahn,giannios}. The fact that this has not been recognized much may be due to the at first sight anti-intuitive nature of the effect, and the emphasis on axisymmetric models in previous work. It may well be the actual mechanism of `Poynting flux conversion' in jets, replacing the mechanisms seen in 2-D \cite{moll09}.

An initially axisymmetric flow of nearly azimuthal magnetic field is bound to be highly unstable to kinking modes. This is to be expected specially when it is highly collimated: in a frame comoving with such a flow, the field is close to a static, almost azimuthal configuration. The details of such configurations are well known since the early days of controlled fusion research. Purely azimuthal fields like this are found to be unconditionally unstable. Instability reduces the energy $B_\phi^2$ of the magnetic field, and small length scales developing under the instability can lead to reconnection, further reducing the magnetic energy. The growth time of the instability is of the order of the Alfv\'en travel time around a loop of azimuthal field. Instability is thus more destructive in highly collimated jets than in wider outflows. Dissipation of magnetic field energy into radiation by such instability has been proposed by \cite{lyutikov} as a mechanism to power the prompt emission of Gamma-ray bursts. 

Another way of dissipating magnetic energy is to generate the flow from a {\em non}-axisym\-metric rotating magnetic field. A classic example is the pulsar wind generated by a rotating neutron star with a  magnetic field inclined with respect to the rotation axis. If the asymmetry is strong enough, the azimuthal field in the outflow will change sign on a length scale $L\sim \pi v/\Omega$  (a `striped wind' \cite{kennel,bogo}), where $\Omega$ is the rotation rate of the source, and $v$ the speed of the outflow. For a relativistic outflow, $L$ is of the order of the light cylinder radius of the rotator. This is generally quite small compared with the distances traveled by the flow. Dissipation of magnetic energy by reconnection of field lines can be very efficient on such short length scales.

The effect of dissipation of magnetic energy on the flow is dramatic. In the absence of dissipation the balance between pressure gradient and curvature force tends to impede the acceleration and the conversion of Poynting flux. In axisymmetry, the balance between these forces can shift in favor of acceleration by the magnetic pressure gradient if (parts of) the flow expand more rapidly than in a radial (`conical') flow (see Section \ref{coneff}). In 3-D, diffusion caused by  instabilities reduces this effect by making the flow more uniform across its crossection. But at the same time, the decay of magnetic energy by instabilities reduces the magnetic pressure along the path of the flow, causing the balance to shift in favor of the pressure gradient (see Eq. \ref{balan}), again resulting in acceleration of the flow. See Section \ref{diverg}  and \cite{moll09}. This effect is efficient in particular in GRB outflows, where observations indicate the largest Lorentz factors \cite{drenkhahn,giannios}.

It may seem strange that one can convert Poynting flux into kinetic energy by `throwing away magnetic energy'. The magnetic energy carried by the flow accounts for only half of the Poynting flux, however. The Poynting flux in MHD is ${\bf v_\perp} B^2/4\pi$, where $\bf v_\perp$ is the velocity component perpendicular to $\bf B$. This is twice the rate of advection of magnetic energy $B^2/8\pi$. The other half is accounted for by the work done by the central engine on the magnetic pressure of the outflow. 

This is entirely analogous to the case of a steady hydrodynamic flow, where the energy balance is expressed by the Bernoulli function. The relevant thermal energy in that case is the enthalpy, the sum of the internal energy (equivalent to $B^2/8\pi$ in our case) and the pressure (also equivalent to $B^2/8\pi$). If internal energy is taken away along the flow (for example by radiation), a pressure gradient develops which accelerates the flow. The energy for this acceleration is accounted for by the $P{\rm d}V$ work done at the source of the flow. For a more extended discussion of this point see \cite{spruit4}.

This mechanism does not require an increasing opening angle of the flow lines. If the dissipation is due to magnetic instabilities, it works best at high degrees of collimation. If it is due to reconnection in an intrinsically nonaxisymmetric flow it works independent of the degree of collimation \cite{drenkhahn}. It may well be one of the main factors determining the asymptotic flow speed in many jets \cite{drenkhahn_spruit,giannios}. However, the mechanism becomes effective mostly at distances significantly beyond the Alfv\'en radius of the flow. Simulations which focus on the region around the black hole, do not usually cover these distances very well (cf. the discussion on length scales in Sect. \ref{len} above). Recent 3-D simulations covering a large range in distance show that dissipation by kink instabilities can effectively become complete after 10-30 Alfv\'en radii \cite{moll,moll09}.

\subsubsection{Observational evidence}
The development of the magnetic acceleration model has raised the question how observations can tell if jets are actually magnetically powered. On the parsec-and-larger scales in AGN jets, there does not appear to be strong evidence for magnetic fields being a major  component of the energy content of the flow \cite{sikora} (for a recent observation see \cite{mehta}). Rather than interpreting this as a failure of magnetic models, it can be understood as evidence for the effectiveness of dissipation of magnetic energy in the flow. Parsec scales in AGN are large compared with the expected Alfv\'en radius in magnetic models, and there is ample opportunity for dissipation by internal instabilities of the helical highly coiled magnetic fields found  in axisymmetric models. 3-D instabilities allow the flow to shed the magnetic field that powered it, so that  it ends as a ballistic, essentially nonmagnetic flow (cf. Section 13.9 in \cite{spruit1}). 

The consequences of this interpretation are significant. First, it implies that there is no point in interpreting magnetic field observations on these large scales in terms of the simple winding-up process happening near the Alfv\'en distance (Fig.\ \ref{windup}). Instabilities will have destroyed the original organized helical fields long before the flow reaches these scales. This questions a popular interpretation of observations of radio polarization in AGN jets. Secondly, flow acceleration is an automatic consequence of internal dissipation of magnetic energy. The effectiveness of such dissipation as indicated by the observations implies that it can be a significant, perhaps even the dominant factor determining the observed jet speeds \cite{giannios}.

\section{Jet collimation}
\label{colli}

In this Section the reader is reminded of the problems with the idea of collimating a jet by magnetic `hoop stress'. The notion that a coiled magnetic field, as in the outflow from a magnetic rotator, will confine itself by hoop stresses is incorrect. Accommodation to this intuition has led to confused discussions in the observational and numerical literature (even where technically correct, e.g. \cite{mckinney2}). Collimation of jets as observed must be due to some external agent; suggestions are discussed at the end of this Section.

To see this, recall that a magnetic field is globally expansive, corresponding to the fact that it represents a positive energy density. That is, a magnetic field can only exist if there is an external agent to take up the stress it exerts. In the laboratory, this agent is a current carrying coil, or the solid state forces in a bar magnet. The stress exterted by the magnetic field on the coils generating them is what limits the strengths of the magnets in fusion devices or particle accelerators. 

A well-known theorem, particularly useful in the astrophysical context is the `vanishing force free field theorem'. A magnetic field on its own, i.e. without other forces in the equation of motion, must be force free, $\bf (\nabla\times B)\times B=0$. The theorem says that if a field is force free everywhere and finite (i.e. vanishing sufficiently fast at infinity), it vanishes identically (for proofs see, e.g. \cite{roberts,mestel,kulsrud}).

In the case of magnetic jets, this means that they can exist only by virtue of a surface that takes up the stress in the magnetic field. In most numerical simulations this is the external medium surrounding the flow. Its presence is assumed as part of the physical model (as in the `magnetic towers') or simply to ease numerical problems with low gas densities. Consider the boundary between the magnetic field (`jet') and the field free region around it in such a calculation. Pressure balance at this boundary is expressed by 
\beq
B^2_{\rm p}+B_\phi^2=P_{\rm e},
\eeq
where ${\bf B}_{\rm p}=(B_r,B_z)$ is the poloidal field and $B_\phi$ the azimuthal component, $P_{\rm e}$ the external pressure, while the internal pressure has been neglected without loss of generality. For a given poloidal field configuration (i.e. shape and magnetic flux of the jet), addition of an azimuthal field component {\em increases} the pressure at the boundary. Everything else being equal, this will cause the jet to {\em expand}, in spite of curvature forces acting in the interior of the jet. 

Confusion about  the role of curvature force is old and reappears regularly in the astrophysical literature.  For a discussion with detailed examples see \cite{parker} (Chap. 9, esp. pp. 170-171). The azimuthal component {\em can} cause constriction to the axis, but only in a {\em part} of the volume, around the axis itself in particular. This part can not live on its own. It is surrounded by a continuation of the field out to a boundary where the stress of the entire configuration can be taken up. These facts are easily recognizable in existing numerical simulations (cf. \cite{uzdensky}).

\subsection{Collimation of large scale relativistic jets}
Since magnetic jets do not collimate themselves, an external agent has to be involved. A constraint can be derived from the observed opening angle $\theta_\infty$. Once the flow speed has a Lorentz factor $\Gamma>1/\theta_\infty$, the different directions in the flow are out of causal contact, and the opening angle does not change any more (at least not until the jet slows down again, for example by interaction with its environment as in the case of a GRB). Turning this around, collimation must have taken place at a distance where the Lorentz factor was still less than $1/\theta_\infty$. 

Once on its way with a narrow opening angle, a relativistic jet needs no external forces to keep it collimated. Relativistic kinematics guarantees that it can just continue ballistically, with unconstrained sideways expansion. This can be seen in a number of different ways. One of them is the causality argument above, alternatively with a Lorentz transformation. In a frame comoving with the jet the sideways expansion is  limited by the maximum sound speed of a relativistic plasma, $c_{\rm s,m}=c/\sqrt 3$. Since it is transversal to the flow, the apparent expansion rate in a lab frame (a frame comoving with the central engine, say) is reduced by a factor $\Gamma$: the time dilatation effect. In the comoving frame, the same effect appears as Lorentz contraction: the jet expands as quickly as it can, but distances to points long its path are reduced by a factor $\Gamma$ (for example the distance to the lobes: the place where jet is stopped by the interstellar medium). In AGN jets, with Lorentz factors of order $\sim 20$, the jet cannot expand to an angle of more than about a degree. This holds if the flow was initially collimated: it still requires that a sufficiently effective collimating agent is present in the region where the jet is accelerated. Comparisons of Lorentz factors and opening angles of AGN jets might provide possible clues on this agent.

\subsection{External collimating agents}

The agent responsible for collimation somehow must be connected with the accretion disk (especially in microquasars where there is essentially nothing else around). One suggestion \cite{bogo2} is that the observed jet is confined by a slower outflow from the accretion disk. In AGN, the `broad line region' outflow might serve this role. Something similar may be the case in protostellar outflows (\cite{santiago} and references therein). In microquasars, such flows are not observed.

Another possibility \cite{spruit3} is that the collimation is due to an ordered magnetic field kept in place by the disk: the field that launches the jet from the center may may be part of a larger field configuration that extends, with declining strength, to larger distances in disk. If the strength of this field scales with the gas pressure in the disk, one finds that the field lines above the disk naturally have a nearly perfectly collimating shape (see analytical examples given by \cite{sakurai87,cao1994}). The presence and absence  of well-defined jets at certain X-ray states would then be related to the details of how ordered magnetic fields are accreted through the disk (cf. Sect. \ref{accord}). 

Near the compact object, the accretion can be in the form of an ion-supported flow (with ion temperatures near virial) which is geometrically thick ($H/r\approx 1$). Jets launched in the central `funnel' of such a disk are confined by the surrounding thick accretion flow. As shown by current numerical simulations, this can lead to a fair degree of collimation, though collimation to angles of a few degrees and less as observed in some sources will probably require an additional mechanism. 

\subsection{Occurrence of instabilities, relation to collimation}
 
 In a cylindrical (i.e. perfectly collimated) jet, the wound-up, azimuthal component of  the field will always be unstable, whether by external or internal kink instabilities. In a rapidly expanding jet, on the other hand,  the Alfv\'en speed drops rapidly with distance, and an Alfv\'enic instability may get `frozen out' before it can develop a significant amplitude.  For which types of collimated jet should we expect instability to be most effective in destroying the azimuthal magnetic field?
 
An estimate can be made by comparing the instability time scale with the expansion time scale of the jet radius. If the jet expands faster than the Alfv\'en speed based on $B_\phi$, there is no time for an Alfv\'enic instability to communicate its information across the jet, and instability will be suppressed.

To see how this works out \cite{moll}, let the distance along the jet be $z$, the jet radius $R(z)$ a function of $z$. As a reference point take the Alfv\'en distance, the distance where the flow speed $v$ first exceeds the Alfv\'en speed  $v_{\rm Ap}$ based on the poloidal field strength (cf. section 1). Call this point $\za$, and denote quantities evaluated at this point with an index $_0$. We then have the approximate equalities:
\beq B_{{\rm p}0}\approx B_{\phi 0}\quad, \qquad v_0\approx v_{{\rm A}0}\quad, \eeq
while the flow speed reaches some modest multiple $k$ of its value at $\za$:
\beq v=k v_0.\eeq
 In the following it is assumed that the jet has reached this constant asymptotic speed $v$.
The shape of the jet $R(z)$ depends on external factors such as an external collimating agent. Assume for the dependence on distance:
\beq R=\epsilon \za({z\over \za})^\alpha. \eeq
I.e. $\epsilon$ is the opening angle of the jet at the Alfv\'en distance.
The mass flux $\dot m$ is constant (steady flow is assumed), so
\beq \dot m=\rho R^2v=\rho_0R_0^2 v=\rho_0\epsilon^2\za^2 v,\eeq
At  $\za$ the
azimuthal field component is of the same order as the poloidal component $B_{\rm p}$. In the 
absence of dissipation by instability,  the azimuthal field strength thus varies with jet radius as
\beq  B_\phi\approx B_{\rm p0}({R\over R_0})^{-1}=B_{\rm p0}({z\over \za})^{-\alpha}. \eeq
The (azimuthal) Alfv\'en frequency is
\beq \omega_{\rm A}(z)=v_{{\rm A}\phi}/R,\eeq
where $v_{{\rm A}\phi}=B_\phi/(4\pi\rho)^{1/2}$. 
The instability rate $\eta$ is some fraction of this:
\beq \eta=\gamma\omega_{\rm A}(z).\eeq
With the expressions for $R$ and $\rho$ this is
\beq \eta=\gamma {v_{\rm A0}\over\epsilon\za}({z\over\za})^{-\alpha}.\eeq
The expansion rate $\omega_{\rm e}$ of the jet is:
\beq \omega_{\rm e}={ {\rm d} \ln R\over{\rm d}t }=v{{\rm d}\ln R\over{\rm d}z}=\alpha{v\over z}.\eeq
The ratio is
\beq
{\eta\over\omega_{\rm e}}={\gamma\over k\epsilon\alpha}({z\over \za})^{1-\alpha}.
\eeq

For an increasingly collimated jet ($\alpha<1$) the instability rate will become larger than the expansion rate at some distance, and kink instability will become important. Decollimating jets ($\alpha>1$)  do not become very unstable since the instability soon `freezes out' due to the decreasing Alfv\'en speed. For the in-between case of a constant opening angle, a conical jet $\alpha=1$, the ratio stays constant and it depends on the combination of factors of order unity $\gamma/(k\epsilon)$ whether instability is to be expected. A numerical study of these effects is given in \cite{moll}.

In unstable cases it may take some distance before the effects of instability become noticeable, depending on the level of perturbations present at the source of the flow. Then again, as noted above (cf. section \ref{len}), in most observed jets the range of length scales is quite large. Even a slowly growing instability can have dramatic effects that do not become evident  in, for example numerical simulations covering a limited range in length scale.

When instability is present, it reduces the azimuthal field strength (since this is what drives the instability) until the growth rate of the instability has settled to a value around the expansion rate $\omega_{\rm e}$. 

Since the decay of magnetic internal energy has an accelerating effect on the flow, a relation between acceleration and collimation is to be expected. Jets which go through an effective (re)collimation stage should achieve a better `Poynting flux conversion' efficiency by dissipation of magnetic energy. This is the opposite of the (axisymmetric) process of acceleration by decollimation discussed in section \ref{diverg}, which yields the best conversion in strongly, in particular nonuniformly, diverging flows.

\section{The launching region}
\label{launch}

As {\em launching region} we define the transition between the high-$\beta$ disk interior and the flow region above the disk. It contains the base of the flow, defined here as the point (called the {\em sonic point}) where the flow speed reaches the slow magnetosonic cusp speed. The mass flux in the jet is determined by the conditions at this point; these are visualized most easily in the centrifugal picture of acceleration. If $\Omega$ is the rotation rate of the footpoint of the field line, $r$ the distance from the axis and $\Phi$ the gravitational potential,  the accelerating force can be derived from an effective potential $\Phi_{\rm e}=\Phi-{1\over 2}\Omega^2 r^2$. As in other hydrodynamic problems, the sonic point lies close to the peak of the potential barrier, the maximum of $\Phi_{\rm e}$. Its height and location depend on $\Omega$ and the strength and inclination of the field. As in the case of hydrodynamic stellar wind theory, the mass flux is then approximately the product of gas density and sound speed at the top of the potential barrier.

\subsection{Models for the disk-flow transition}
If the Alfv\'en surface is not very close to the disk surface, the magnetic field in the disk atmosphere is approximately force free since the gas pressure declines rapidly with height. As with any force free or potential field, the shape of the field lines in this region is a {\em global} problem. The field at any point above the disk is determined by the balance of forces inside the field: field lines sense the pressure of their neighbors. At points where the field strength at the underlying disk surface is high, the field lines above it spread away from each other, like the field lines at the pole of a bar magnet. 

The inclination of the field lines at the base of the flow is thus determined in a global way by the distribution of field lines at the disk surface, i.e. the vertical component $B_z(r)$ (assuming axisymmetry for this argument). Most of the physics inside a thin disk can be treated by a local approximation, that is, only a region with a radial extent similar to the disk thickness needs to be considered. The field inclination  at the base of the flow, however,  a key factor in the launching problem, cannot be computed in this way.  

Several more mathematically inclined studies have nevertheless attempted to find `selfconsistent' field configurations  in a local approximation,  by extrapolating field configurations along individual field lines from inside the disk into the flow region  (\cite{wardle}, \cite{ferreira} and references therein, \cite{shalybkov,campbell}). By ignoring magnetic forces in the low-$\beta$ region, these results do not yield the correct field configuration above the disk except in singular cases. The high-$\beta$ disk interior and the low-$\beta$ disk atmosphere are regions of different physics, and so are the factors determining the field line shape in these regions. 

The transition from the disk to the flow regime can still be studied in a local approximation provided one gives up the ambition of at the same time determining the field configuration above the disk. Since the field inclination is determined also by conditions at distances that are not part of the local region studied,  the inclination at some height above the disk then has to be kept as an {\em external parameter}  in such a local study. 

This has been done in the detailed study by \cite{ogilvie}. Their results show how the mass flux depends on the strength of the field and its asymptotic inclination.  If the magnetic stress $f_r=B_zB_r/4\pi$ is kept fixed, the mass flow increases with increasing inclination of the field lines with respect to the vertical as expected. The flow rate  {\em decreases} with increasing field strength, however. This is due to the fact that the curvature of magnetic field lines shaped as in Fig.\ \ref{regions} exerts the outward force $f_r$ (against gravity) on the disk. The rotation rate is therefore a bit lower than the Keplerian value. This is equivalent to an increase of the potential barrier in $\Phi_{\rm e}$ for mass leaving the disk along field lines. 

This complicates the conditions for launching a flow from the disk, compared with the simple estimate based on the field line inclination alone.  Ignoring the slight deviation from Kepler rotation, a cool disk would launch a flow only if the inclination of the field lines with respect to the vertical is greater than $60^\circ$ \cite{blandfordpayne}. This condition is significantly modified by the subkeplerian rotation of the field lines, especially for the high magnetic field strengths that may be the most relevant for the generation of jets.

\subsubsection{Limitations in numerical simulation}
\label{diffic}
For a convincing numerical treatment of the launching and initial acceleration of the flow, the calculations would have to cover both the high-$\beta$ disk interior and the low-$\beta$ atmosphere. This implies a large range in characteristic velocities to be covered, with the region of large Alfv\'en speeds limiting the time step. To circumvent this {\em time scale problem}, modifications of the MHD equations have been explored in which the speed of magnetic waves is artificially reduced, cf.\ \cite{miller}. The problem is alleviated somewhat in relativistic calculations of  flows near a black hole, where the various characteristic velocities of the problem converge on the speed of light.

Another way to reduce the time scale problem is by choosing conditions corresponding to a high mass flux, thus increasing the density and decreasing the Alfv\'en speeds in the wind. The Alfv\'en surface then decreases in size. This has the added benefit that  the conceptually different regions in the flow (cf. Fig.\ \ref{regions}) fit more comfortably inside an affordable computational volume . 

It has to be realized, however, that this choice also limits the results to a specific corner of parameter space that may or may not be the relevant one for observed jets. It limits the jet speeds reached since the asymptotic flow speed decreases with mass loading. It strongly limits the generality of quantitative conclusions (in particular, about the mass flux in the jet, cf.\  \cite{keppens}). It also tends to bias interpretations of jet physics to ones that are most meaningful in the high mass flux corner of parameter space (cf. discussion in Sect. \ref{hiflo}).

\subsubsection{Ion supported flows}
At low disk temperature, the conditions for outflow are sensitive to the field strength and inclination near the disk surface, raising the question why the right conditions would be satisfied in any given jet-producing object. This sensitivity is much less in an ion-supported flow \cite{rees}, where the (ion-)temperature of the flow is near the virial temperature and the flow only weakly bound in the gravitational potential of the accreting object. This may, in part, be the reason why powerful jets tend to be associated with the hard states in X-ray binaries for which the ion-supported flow (also called ADAF \cite{chen}) is a promising model.

\subsection{Instability of the disk-wind connection, knots}
The same sensitivity of the mass flux to configuration and strength of the field can cause the connection between disk and outflow to become {\em unstable}. Since the wind carries angular momentum with it, an increase in  mass loss in the wind causes an increase in the inward  drift speed of the disk at the footpoints of the flow. This drift carries the vertical component of the magnetic field with it. Since the field configuration in the wind zone is determined by the distribution of footpoints on the disk, this feeds back on the wind properties.  Linear stability analysis by \cite{cao2002} showed that this feedback leads to inward propagating unstable disturbances, with associated variations in mass flux in the wind. This had been assumed before in a model by  \cite{aga}. These authors found that this kind of feedback causes the disk to become dramatically time dependent in a manner suggestive of the  FU Ori outbursts in protostellar disks. 

Strong ordered magnetic fields in disks have their own forms of instability,  independent of the coupling to a wind \cite{lubow2,spruit2}, driven instead by the energy in the field itself. The nonlinear evolution of such instabilities was studied numerically by \cite{stehle} (see also Sect.\ \ref{strength} above). Their effect appeared to be similar to an enhancement of the rate of diffusion of the magnetic field through the disk. 

Since both these kinds of instability cause changes in the vertical component of the field, which is the same on both side of the disk, they produce symmetric variations in mass flow in jet and counterjet. They are thus good candidates for the time-dependence often observed in the form of symmetric knot patterns in protostellar jets.


\begin{thebibliography}{99}

\bibitem{aga} V.~Agapitou: Ph.D.~Thesis,  Queen Mary and Westfield College (2007)

\bibitem{anderson} J.M.~Anderson, Z.-Y.~Li, R.~Krasnopolsky, R.D.~Blandford: \apj \textbf{630}, 945 (2005)

\bibitem{bains} I.~Bains, A.M.A.~Richards, T.M.~Gledhill et al.: \mnras \textbf{354}, 529 (2004) 

\bibitem{barkov} M.~Barkov, S.~Komissarov: arXiv:0801.4861v1 [astro-ph] (2008)

\bibitem{beckwith}K.~Beckwith, J.~F.\ Hawley, \& J.~H.\ Krolik: \apj \textbf{678}, 1180 (2008)

\bibitem{begelman} M.C.~Begelman, Z.-Y.~Li: \apj \textbf{426}, 269 (1994)

\bibitem{belloni05} T.~Belloni, J.~Homan, P.~Casella et al.: A\&A, \textbf{440}, 207 (2005)

\bibitem{kogan74} G.S.~Bisnovatyi-Kogan, A.A.~Ruzmaikin: \apss \textbf{28}, 45 (1974)

\bibitem{kogan76} G.S.~Bisnovatyi-Kogan, A.A.~Ruzmaikin: \apss \textbf{42}, 401 (1976)

\bibitem{BZ}  R.~D.~Blandford, \& R.~L.\ Znajek:  \mnras \textbf{179}, 433 (1977) 

\bibitem{blandfordpayne} R.D.~Blandford, D.G.~Payne: \mnras \textbf{199}, 883 (1982)

\bibitem{bogo} S.V.~Bogovalov: \aap \textbf{349}, 1017 (1999)

\bibitem{bogo2} S.V.~Bogovalov, K.~Tsinganos: \mnras \textbf{357}, 918 (2005)

\bibitem{campbell} C.G.~Campbell: \mnras \textbf{345}, 123 (2003)

\bibitem{cao1994} X.~Cao, H.C.~Spruit: \aap \textbf{287}, 80 (1994)

\bibitem{cao2002}  X.~Cao, H.C.~Spruit: \aap \textbf{385}, 289 (2002)

\bibitem{keppens} F.\ Casse, \& R.\ Keppens: \apj \textbf{601}, 90 (2004)

\bibitem{chen}  X.~Chen, M.A.~Abramowicz, J-.P.~Lasota et al: \apjl \textbf{443}, L61 (1995)

\bibitem{churazov} E.~Churazov, M.~Gilfanov, M.~Revnivtsev: \mnras \textbf{321}, 759 (2001)

\bibitem{daigne} F.~Daigne,  G.~Drenkhahn: \aap \textbf{381}, 1066 (2002)

\bibitem{dangelo}C.~D'Angelo, D.~Giannios,  C.~Dullemond, Spruit : , \aap 488, 441(2008)

\bibitem{villiers} J.-P.~De Villiers, J.F.~Hawley, J.H.~Krolik et al.: \apj \textbf{620}, 878 (2005)

\bibitem{drenkhahn} G.~Drenkhahn: \aap \textbf{387}, 714 (2002)

\bibitem{drenkhahn_spruit} G.~Drenkhahn, H.C.~Spruit: \aap \textbf{391}, 1141 (2002)

\bibitem{fabian} A.C.~Fabian, P.W.~Guilbert, C.~Motch et al.:  \aap \textbf{111}, L9 (1982)

\bibitem{ferreira} J.~Ferreira, G.~Pelletier: \aap \textbf{295}, 807 (1995)

\bibitem{fromang} S.~Fromang, J.C.B.~Papaloizou, G.~Lesur et al.: \aap, \textbf{476}, 1123  (2007)

\bibitem{giannios} D.~Giannios, H.C.~Spruit: \aap \textbf{450}, 887 (2006)

\bibitem{goldreich} P.~Goldreich, W.H.~Julian:  \apj \textbf{160}, 971 (1970)

\bibitem{guan} X. Guan, C.~F. Gammie, J.~B. Simon, B.~M. Johnson: 2009, \apj \textbf{694}, 1010 

\bibitem{hawley} J.F.~Hawley, J.H.~Krolik: \apj \textbf{641}, 103 (2006)

\bibitem{heinemann} M.~Heinemann, S.~Olbert:  \jgr \textbf{83}, 2457, (erratum in  \jgr \textbf{84}, 2142) (1978)

\bibitem{HomBel05} J.~Homan, T.~Belloni: Ap\&SS, \textbf{300}, 107 (2005)

\bibitem{homan} J.~Homan VI$^{\rm th}$ Microquasar Workshop: Microquasars and Beyond, 
Proceedings of Science (http://pos.sissa.it), PoS(MQW6)093 (2006)

\bibitem{hutawara} B.~Hutawarakorn, T.J.~Cohen, G.C.~Brebner:  \mnras \textbf{330}, 349 (2002)

\bibitem{igu} I.V.~Igumenshchev, R.~Narayan, M.A.~Abramowicz: \apj \textbf{592}, 1042 (2003)

\bibitem{kanbach} G.~Kanbach, C.~Straubmeier, H.C.~Spruit et al.:  Nature \textbf{414}, 180 (2001)

\bibitem{kennel} C.F.~Kennel, F.V.~Coroniti:  \apj \textbf{283}, 710 (1984)

\bibitem{komissarov} S.~S.\  Komissarov: \  arXiv:0804.1912 (2008)

\bibitem{krolik} J.H.~Krolik, J.~Hawley: VI$^{\rm th}$ Microquasar Workshop: Microquasars and Beyond, Proceedings of Science (http://pos.sissa.it), PoS(MQW6)046  (2006)

\bibitem{kulsrud} R.~Kulsrud, R.: Plasma Physics for Astrophysics, Princeton University Press (2006)

\bibitem{lubow1} S.H.~Lubow, J.C.B.~Papaloizou, J.E.~Pringle: \mnras \textbf{267}, 235 (1994)

\bibitem{lubow2} S.H.~Lubow, H.C.~Spruit: \apj \textbf{445}, 337 (1995)

\bibitem{lyutikov} M.~Lyutikov, R.~Blandford: arXiv:astro-ph/0312347 (2003)

\bibitem{machidaa} M.~Machida,  M.R.~Hayashi, R.~Matsumoto: ApJ \textbf{532}, L67 (2000)

\bibitem{machidab} M.~Machida, R.~Matsumoto, S.~Mineshige: ArXiv e-prints, arXiv:astro-ph/0009004v1 (2000)

\bibitem{matsumoto} R.~Matsumoto, T.~Matsuzaki, T.~Tajima et al: Astrophysics and Space Science Library  \textbf{263}, 247 (2001)

\bibitem{mckinney1} J.C.~McKinney, C.F.~Gammie:  \apj \textbf{611}, 977 (2004)

\bibitem{mckinney2} J.C.~McKinney, R.~Narayan: arXiv:astro-ph/0607575v1 (2006)

\bibitem{mckinney3} J.C.~McKinney, R.~Narayan: \mnras \textbf{375}, 513 (2007)

\bibitem{mehta} K.~T. Mehta, M.~Georganopoulos., E.~S.~Perlman, C.~A.~Padgett,
G.~Chartas:  \apj \textbf{690}, 1706 (2009)

\bibitem{mestel} L.~Mestel: Stellar magnetism, (Clarendon, Oxford,1999), (International series of monographs on physics) (1999)

\bibitem{michel} F.C.~Michel: \apj \textbf{158}, 727 (1969)

\bibitem{miller} K.A.~Miller, J.M.~Stone: \apj \textbf{534}, 398 (2000)

\bibitem{moll} R.~Moll, H.C.~Spruit, M.~Obergaulinger, \aap \textbf{492}, 621 (2008)

\bibitem{moll09} R.~Moll \aap submitted (2009)

\bibitem{ogilvie} G.I.~Ogilvie, M.~Livio: \apj \textbf{553}, 158 (2001)

\bibitem{parker} E.N.~Parker: Cosmical Magnetic Fields, Clarendon Press, Oxford (1979)

\bibitem{phinney} E.S.~Phinney: Ph.D.~Thesis,  University of Cambridge, UK (1983)

\bibitem{rees} M.J.~Rees, M.C.~Begelman, R.D.~Blandford et al.: \nat \textbf{295}, 17 (1982)

\bibitem{roberts} P.H.~Roberts: Magnetohydrodynamics, Longmans, London, Ch 4.4 (1967)


\bibitem{rutledge} R.E.~Rutledge et al.: ApJS \textbf{124}, 265 (1999)

\bibitem{sakurai85} T.~Sakurai: \aap \textbf{152}, 121 (1985)

\bibitem{sakurai87} T.~Sakurai: \pasj \textbf{39}, 821 (1987)

\bibitem{santiago} J.~Santiago-Garc\'{\i}a, M.~Tafalla, D.~Johnstone, R.~Bachiller: ArXiv e-prints, arXiv:0810.2790 (2008)

\bibitem{shalybkov} D.~Shalybkov, G.~R\"udiger: \mnras \textbf{315}, 762 (2000)

\bibitem{sikora} M.~Sikora, M.~C.~ Begelman, 
G.~M.~Madejski, J.-P.~Lasota: \apj \textbf{625}, 72 (2005)

\bibitem{spencer}  R.~Spencer: in VI$^{\rm th}$ Microquasar Workshop: Microquasars and Beyond, Proceedings of Science (http://pos.sissa.it), PoS(MQW6)053 (2006)

\bibitem{spruit1} H.C.~Spruit: NATO ASI Proc.\  C477: Evolutionary Processes in Binary Stars, p249 (arXiv:astro-ph/9602022v1) (1996)

\bibitem{spruit2} H.C.~Spruit, R.~Stehle, J.C.B.~Papaloizou: \mnras \textbf{275}, 1223 (1995)

\bibitem{spruit3} H.C.~Spruit, T.~Foglizzo, R.~Stehle: \mnras \textbf{288}, 333 (1997)

\bibitem{spruit4} H.C.~Spruit, G.D.~Drenkhahn: Astronomical Society of the Pacific Conference Series, 312, 357 (2004)

\bibitem{spruit5} H.C.~Spruit, D.A.~Uzdensky: \apj \textbf{629}, 960 (2005)

\bibitem{spruit6} H.C.~Spruit:  in VI$^{\rm th}$ Microquasar Workshop: Microquasars and Beyond, Proceedings of Science (http://pos.sissa.it), PoS(MQW6)044 (2006)

\bibitem{stehle} R.~Stehle, H.C.~Spruit: \mnras \textbf{323}, 587 (2001)

\bibitem{tchek} A.\ Tchekhovskoy,  J.~C.\ McKinney, R.\ Narayan:  arXiv:0901.4776 (2009)

\bibitem{uzdensky} D.A.~Uzdensky, A.I.~MacFadyen:  \apj \textbf{647}, 1192 (2006)

\bibitem{vanballegooijen} A.A.~van Ballegooijen: Astrophysics and Space Science Library, Kluwer, \textbf{156}, 99 (1989)

\bibitem{vanderklis} M.~van der Klis: \apj \textbf{561}, 943 (2001)

\bibitem{vlemmings} W.H.T.~Vlemmings, H.J.~van Langevelde, P.J.~Diamond:\aap \textbf{434}, 1029 (2005)

\bibitem{wardle} M.~Wardle, A.~K\"onigl: \apj \textbf{410}, 218 (1993)

\end{thebibliography}
\end{document}